\documentclass{article}

\usepackage{arxiv}

\usepackage[utf8]{inputenc} 
\usepackage[T1]{fontenc}    
\usepackage{hyperref}       
\usepackage{url}            
\usepackage{booktabs}       
\usepackage{amsfonts}       
\usepackage{nicefrac}       
\usepackage{microtype}      
\usepackage{amsmath}
\usepackage{lipsum}
\usepackage{authblk}
\usepackage{lmodern}
\usepackage{graphicx}
\graphicspath{ {./images/} }

\title{Benchmarking atmospheric circulation variability in an AI emulator, ACE2, and a hybrid model, NeuralGCM}

\author[1]{\textbf{I. Baxter\thanks{Corresponding author: itbaxter@uchicago.edu}}}
\author[1,2]{\textbf{H. A. Pahlavan}}
\author[1]{\textbf{P. Hassanzadeh}}
\author[1]{\textbf{K. Rucker}}
\author[1]{\textbf{T. A. Shaw}}

\affil[1]{Department of the Geophysical Sciences, The University of Chicago, Chicago, IL}
\affil[2]{Northwest Research Associates, Boulder, CO}

\begin{document}
\maketitle



%
%

%
%

\begin{abstract}
Physics-based atmosphere-land models with prescribed sea surface temperature have notable successes but also biases in their ability to represent atmospheric variability compared to observations. Recently, AI emulators and hybrid models have emerged with the potential to overcome these biases, but still require systematic evaluation against metrics grounded in fundamental atmospheric dynamics. Here, we evaluate the representation of four atmospheric variability benchmarking metrics in a fully data-driven AI emulator (ACE2-ERA5) and hybrid model (NeuralGCM). The hybrid model and emulator can capture the spectra of large-scale tropical waves and extratropical eddy-mean flow interactions, including critical levels. However, both struggle to capture the timescales associated with quasi-biennial oscillation (QBO, $\sim 28$ months) and Southern annular mode propagation ($\sim 150$ days). These dynamical metrics serve as an initial benchmarking tool to inform AI model development and understand their limitations, which may be essential for out-of-distribution applications (e.g., extrapolating to unseen climates).
\end{abstract}


\section*{Keypoints}
\begin{itemize}
    \item Atmospheric circulation variability in an AI emulator and hybrid model is benchmarked against ERA5 and physics-based models. 
    \item AI models successfully capture tropical wave variability and extratropical wave-mean flow interactions.
    \item AI models struggle to capture quasi-periodic, longer timescale oscillations, such as QBO and propagating SAM.
\end{itemize}

\section*{Plain Language Summary}
The second revolution of AI weather emulators is now being extended to longer timescale atmosphere-land emulation forced by observed sea surface temperatures. We present a set of four benchmarking metrics to evaluate the skill of the currently available AI-physics hybrid model (NeuralGCM) and fully data-driven emulator (ACE2-ERA5) in representing important atmospheric processes against observations and physics-based models. These metrics examine the representation of interactions between waves and the background wind fields in the stratosphere, tropics, and extratropics. We find that AI emulators and hybrid models perform comparably to or better than physics-based models in representing physical processes with timescales of days to weeks, comparable to what the AI models have learned (6 hours). However, both physics-based and AI models struggle with important phenomena with slower timescales of months and years. These results have implications for the current AI models ability to capture climate variability across scales.

%
%

\section{Introduction}

Weather forecasting has recently gone through a second revolution with the emergence of skillful AI weather ``emulators'' \cite{pathakFourCastNetGlobalDatadriven2022,keislerForecastingGlobalWeather2022,biAccurateMediumrangeGlobal2023,lam2023learning}. These neural network-based fully data-driven models are often trained on $\sim$40 years of ERA5 reanalysis, learning to autoregressively predict the atmospheric state—typically at 6-hour intervals. With tens or hundreds of millions of trainable parameters, they infer the statistical relationships that govern atmospheric evolution directly from observation-derived data. More recently, hybrid models such as NeuralGCM \cite{kochkovNeuralGeneralCirculation2024}, which use the dynamical core of the physics-based models but represent all subgridscale physics with a neural network, have emerged too. In this ``hybrid'' model, the dynamical core is differentiable, which allows the  neural network to be trained such that the evolution of the atmospheric state matches that of ERA5 for multiple time steps, gradually extended from 6 hours to 5 days during training. For short- and medium-range forecasting,  the newest AI weather emulators and NeuralGCM can robustly surpass the skills of the best physics-based numerical weather prediction models while operating orders of magnitude faster \cite{kochkovNeuralGeneralCirculation2024,lang2024aifs, price2025probabilistic,ben2024rise,bracco2025machine}. 

Building on this AI revolution for weather forecasting, an initial-value problem, AI-based models are emerging that are focused on subseasonal-to-seasonal and longer-term climate predictions, a boundary-value problem. The AI models, emulator and hybrid, are trained based on supervised learning of fast dynamics (Eq.~(\ref{eq:loss})). The evaluations in this study are meant to quantitatively test the AI models' ability to capture complex dynamics of key atmospheric processes, well beyond what is explicitly in this loss function, and what common metrics such as pattern correlation or root-mean-square error assess.

There are currently two models that are trained to emulate ERA5 over the satellite era (1980 to present). One is a hybrid AI-physics model, NeuralGCM \cite{kochkovNeuralGeneralCirculation2024}, and the other one is an AI atmospheric emulator, ACE2-ERA5 \cite{watt-meyerACE2AccuratelyLearning2025a}, each with prescribed boundary forcings. Both the AI emulator and hybrid model have been shown to outperform physics-based models in several popular climate metrics, such as the number of tropical storms and trends in mean temperature profiles, at a fraction of the computational cost.

Given the rapid rise of AI-based models for longer-term atmosphere (months to years) and even climate predictions \cite{dheeshjithSamudraAIGlobal2025}, and their increasing use for out-of-sample \cite{mengDeepLearningAtmospheric2025} and even out-of-distribution predictions in the future, it is critical to rigorously evaluate these models with fundamentals of atmospheric dynamics. This benchmarking is an important step toward establishing these AI models' skill and trustworthiness, e.g., ensuring that they are getting the right answers for the right reasons. Such tests for AI weather emulators have started to appear, for example, showing their challenges with capturing the butterfly effect and hydrostatic and cyclogeostrophic balances as well as forecasting the rarest weather extremes \cite{bonavitaLimitationsCurrentMachine2024,selz2023can,sunCanAIWeather2025,sun2025predicting}. However, the AI-based models for longer-term predictions have not been studied as much, beyond recent Green's function experiments with ACE2 \cite{kentSkilfulGlobalSeasonal2025,chienModulationTropicalCyclogenesis2025, vanloonReanalysisBasedGlobalRadiative2025} where there are no observations to compare to.

Here, we focus on benchmarking atmospheric circulation variability, which  shapes surface climate and is strongly linked to predictability across timescales. For example, baroclinic instability, which drives weather in the extratropics, contributes to error growth and limits weather prediction \cite{craigWavesWeatherExploring2021}. Seasonal predictability is tied to longer timescale atmospheric variability in the tropics (e.g., Madden Julian Oscillation or MJO, quasi-biennial oscillation or QBO) and extratropics (e.g., baroclinic annular mode, propagating annular mode, stratospheric sudden warmings) \cite{kidsonIndicesSouthernHemisphere1988, sonPreferredModesVariability2006, thompsonPeriodicVariabilityLargeScale2014, vitartSubseasonalSeasonalPrediction2018, sheshadriPropagatingAnnularModes2017}. Specifically, the primary modes are QBO and MJO are in the tropical atmosphere and the annular modes in the midlatitudes, with each having important links to subseasonal-to-seasonal forecast skills \cite{ansteyImpactsProcessesProjections2022}. 

Capturing atmospheric circulation variability is extremely complex because it involves the dynamics of waves and eddies as they are generated via diabatic processes and interact with the time- or zonal-mean flow. These processes involve multi-scale interactions from planetary-scale to turbulent dissipation and can be quasi-linear to fully non-linear across time and spatial scales \cite{vallisEssentialsAtmosphericOceanic2019}. Thus, simulating atmospheric circulation variability is an important dynamical test of the coupling across spatio-temporal scale coupling. 

The current standard for accurately simulating atmospheric circulation variability are physics-based atmosphere-land models. The standard configuration couples an atmospheric general circulation model with parameterizations of subgrid scale physics (radiation, clouds, convection) to a prognostic land model, while prescribing sea surface temperatures (SST) and sea ice concentrations from observations. The Atmosphere Model Intecomparison Project (AMIP) represents a systematic benchmarking of such atmosphere-land models and their ability to capture both climatological features along with atmospheric variability \cite{gatesAMIPAtmosphericModel1992, eyringOverviewCoupledModel2016}. While successive generations of models have improved their representation of atmospheric variability, notable discrepancies remain, particularly in simulating QBO and annular mode characteristics \cite{scaifeSignaltonoiseParadoxClimate2018, orbeRepresentationModesVariability2020, bushellEvaluationQuasiBiennialOscillation2022, lubisIntrinsic150DayPeriodicity2023}. 

Here, we benchmark 4 key features of large-scale atmospheric variability: 
\begin{enumerate}
\item QBO, which is primarily driven by tropospheric gravity waves, with additional contributions from planetary-scale equatorial waves, through their interaction with the stratospheric mean flow,
\item convectively coupled equatorial waves, representing the organization and interaction of tropical rainfall with the mean flow,
\item extratropical synoptic eddy-mean flow interaction, involving wave breaking and formation of critical levels,
\item poleward-propagation of the Southern Annular Mode (SAM), which results from the interaction of synoptic eddies and low-frequency zonal wind variability.
\end{enumerate}
 These features in ERA5 are compared to those in the AI emulator, hybrid model and physics-based atmosphere-land AMIP models participating in the Coupled Model Intercomparison Phase project 6 (CMIP6) \cite{eyringOverviewCoupledModel2016}. These four benchmarks are chosen because they span a broad range of tropical and extratropical variability and because they can be calculated from the current AI emulator and hybrid models' output. They provide a quantitative benchmark of the AI-based models' representation of atmospheric dynamics.

\section{Data and Methods}

\subsection{Models}
The physics-based land-atmosphere models used in this study are from the AMIP configuration available in CMIP6 \cite{gatesAMIPAtmosphericModel1992, eyringOverviewCoupledModel2016}. From these simulations, we use daily precipitation and horizontal winds ($u$, $v$) on standard pressure levels from 1981 to 2014. A summary of the models, realizations, and variables are available in Table S1.

The publicly available NeuralGCM~\cite{kochkovNeuralGeneralCirculation2024} and ACE2-ERA5~\cite{watt-meyerACE2AccuratelyLearning2025a} were also used in this study. Both models are forced with ERA5 SST and sea ice concentrations, while ACE2-ERA5 additionally includes global-mean CO$_2$ concentration as a forcing. Both are initialized at lagged time steps throughout 1980 (detailed below), therefore our evaluation focuses on the AMIP historical period (1981 to 2014). The evaluation period includes the training set for both NeuralGCM and ACE2-ERA5, but we find the diagnostics are robust across training and testing periods. We test the sensitivity of each benchmarking metric over the full and testing periods of the AI emulator and hybrid model, and do not find differences in the modes of variability captured over each time period.

ACE2-ERA5~\cite{watt-meyerACE2AccuratelyLearning2025a} is a fully data-driven atmospheric emulator from the Allen Institute that evolves the atmospheric state $x(t)$ every $\Delta t=6$ hours: ${x}(t+\Delta t)=N(x(t),b(t),\theta)$, where the weights $\theta$ are determined by minimizing loss function
\begin{equation}
\label{eq:loss}
\mathcal{L}= \| {x}(t+\Delta t) - N(x(t),b(t),\theta) \|_2.
\end{equation}
$N$ is a deep neural network based on the Spherical Fourier Neural Operator (SFNO) architecture. The emulator, trained on short 6-hour timesteps, can autoregressively produce forecasts that are indefinitely stable. During training, ${x}(t)$ and ${x}(t+\Delta t)$ in Eq.~\eqref{eq:loss} are from ERA5 (on 1.0$^{\circ}$ horizontal resolution and 8 terrain-following vertical layers) over multiple periods covering 1940–1995, 2011–2019, 2021–2022. $b(t)$ in Eq.~\ref{eq:loss} represents the boundary forcings, which includes ERA5 sea surface temperatures, sea ice concentrations, downward solar radiation and, in the case of ACE2-ERA5, carbon dioxide concentration. A validation period from 1996--2000 is used to select the checkpoint with the best performance (global root mean square error) in 5-year means \cite{watt-meyerACE2AccuratelyLearning2025a}.

In this study, ACE2-ERA5 horizontal winds ($u$ and $v$) are remapped from hybrid-sigma to the same standard pressure levels available from ERA5 and NeuralGCM. We also use precipitation rate (PRESsfc) when examining convectively coupled waves. A 37-member lagged ensemble is initialized with ERA5 data on every 10th day of 1980, and integrated from 1980-2022. All 37 ensemble members were stable for the entire integration.

NeuralGCM is a hybrid atmospheric model composed of a differentiable dynamical core for solving the discretized governing dynamical equations and a single learned physics module that parameterizes all subgridscale physics using a neural network~\cite{kochkovNeuralGeneralCirculation2024}. The model is trained over the period 1979--2018 using a weighted sum of three loss functions similar to Eq.~(\ref{eq:loss}), designed to promote accuracy and sharpness. NeuralGCM is trained exclusively on 6 hour to 5 day rollouts~\cite{kochkovNeuralGeneralCirculation2024}. We inference 37 ensemble members of NeuralGCM using the same initial conditions described earlier. We use the $\mathrm{2.8^{\circ}}$ horizontal resolution deterministic model with 37 standard pressure levels from NeuralGCM that holds global mean surface pressure fixed~\cite{kochkovNeuralGeneralCirculation2024}, ensuring all 37 ensemble members are stable for the entire simulation (1980 to 2023). From the hybrid model, we use horizontal winds ($u$ and $v$) and precipitation minus evaporation ($P\ minus\ E$) in the place of precipitation.

\subsection{Diagnostics}
All outputs throughout this study are first regridded using conservative interpolation to the same 2.8 $\mathrm{^{\circ}}$ horizontal grid used by NeuralGCM. All variables are resampled to daily means for comparison with AMIP models. 

\subsubsection{Quasi-biennial oscillation (QBO)}
The QBO, is characterized by the downward propagation of successive westerly and easterly winds with an average period of $\mathrm{\sim{28}}$ months \cite{baldwinQuasibiennialOscillation2001,ansteyImpactsProcessesProjections2022}. In this study, the QBO index is defined as the monthly and latitude-weighted (10$\mathrm{^{\circ}S}$ to 10$\mathrm{^{\circ}N}$) mean zonal winds at 50 hPa. In addition to ERA5, we show the QBO index from individual ensemble members from the AMIP, ACE2-ERA5, and NeuralGCM ensembles, with the full ensembles in Figure~\ref{fig:Sup. Figure 1}). The displayed members in Figure \ref{fig:Figure 1}a are chosen as representative of the spread in amplitude and periodicity of the QBO across each ensemble. The maximum amplitudes of the QBO index are taken as the difference between the maximum and minimum monthly mean zonal winds at 50 hPa. The periodicity of the QBO is estimated as the peak in the power spectrum of the monthly mean zonal winds at 50 hPa from 1980-2014.

\subsubsection{Equatorial wavenumber-frequency spectra}
In the tropics, atmospheric circulation and its associated impacts are generally determined by atmospheric waves coupled to deep convection \cite{kiladisConvectivelyCoupledEquatorial2009}; however, this convection that is normally parameterized in physics-based models (often poorly) is implicitly learned by the neural networks in the AI emulator and hybrid model. Specific regions of the wavenumber-frequency spectra have been associated with leading modes of tropical variability, such as the MJO and equatorial Rossby, Kelvin, Mixed Rossby-gravity (MRG), and inertio-gravity (IG) waves. To test the representation of convectively coupled waves, we compute frequency-wavenumber spectra following Wheeler and Kiladis (1999)~\cite{wheelerConvectivelyCoupledEquatorial1999}, using daily precipitation from ERA5, AMIP, and ACE2-ERA5. This version of NeuralGCM does not prognostically forecast precipitation directly, but it is inferred from total column moisture convergence (i.e., precipitation minus evaporation). Area-weighted daily mean precipitation rate (or total column moisture convergence) is decomposed into background, symmetric, and asymmetric components about the equator then averaged from $15^{\circ}S$ to $15^{\circ}N$.

\subsubsection{Eddy momentum flux co-spectra}
Extratropical atmospheric dynamics is dominated by the interactions of eddies (deviations from the zonal mean) and the jet stream (mean flow). This interaction can be quantified by eddy flux co-spectra, which are here computed following previous work \cite{hayashiGeneralizedMethodResolving1971, randelPhaseSpeedSpectra1991,chenPhaseSpeedSpectra2007, lutskoLowerTroposphericEddyMomentum2017}. DJFM and JJAS mean 250 hPa eddy momentum flux are calculated from anomalous daily mean horizontal winds ($u$, $v$) from ERA5, AMIP, ACE2-ERA5, and NeuralGCM following the approach of \cite{randelPhaseSpeedSpectra1991}. Daily mean anomalies are computed by removing the annual cycle and tapered by a Hanning window. For each year, $u$ and $v$ are transformed from time and space dimensions to frequency and wavenumber space, then summed over all wavenumbers. The spectra are linearly interpolated to angular phase speed bins and smoothed using a Gaussian filter. The 1981-2014 climatology of the eddy flux cospectra and zonal mean winds are plotted as a function of latitude and angular phase speed. 

\subsubsection{The propagating of the Southern Annular Mode}
The large-scale circulation in the Southern Hemisphere is dominated by the Southern Annular Mode (SAM), which has a 150-day periodicity \cite{lubisIntrinsic150DayPeriodicity2023}. The timescale of the annular mode is determined by wave-mean flow interactions in the extratropics. Eddy feedbacks and their timescales has been used to evaluate the variability of physics-based models \cite{gerberAnnularModeTime2008}. Here, we follow the approach in Lubis and Hassanzadeh (2023)~\cite{lubisIntrinsic150DayPeriodicity2023} to compute the propagation of SAM, using daily mean horizontal wind data at 500 hPa from ERA5, AMIP, ACE2-ERA5, and NeuralGCM from January 1981 to December 2014. 

Anomalies are computed by removing the mean seasonal cycle, defined as the annual average and the first four Fourier harmonics of the daily climatology at each grid point \cite{lorenzEddyZonalFlow2001}. Before calculating the empirical orthogonal functions (EOFs), the vertically averaged, zonal-mean zonal wind anomalies from 20° to 80°S are calculated. Using 500 hPa winds rather than the vertical average, taken from the surface up to 200 hPa, does not change the results. The fields are weighted by the square root of the cosine of the latitude to account for the decrease in area toward the pole prior to EOF analysis \cite{lubisIntrinsic150DayPeriodicity2023,simpsonSouthernAnnularMode2013}. 

 \section{Results} 

\subsection{Quasi-biennial oscillation (QBO)}
\begin{figure}[h!]
    \centering
    \includegraphics[width=\textwidth]{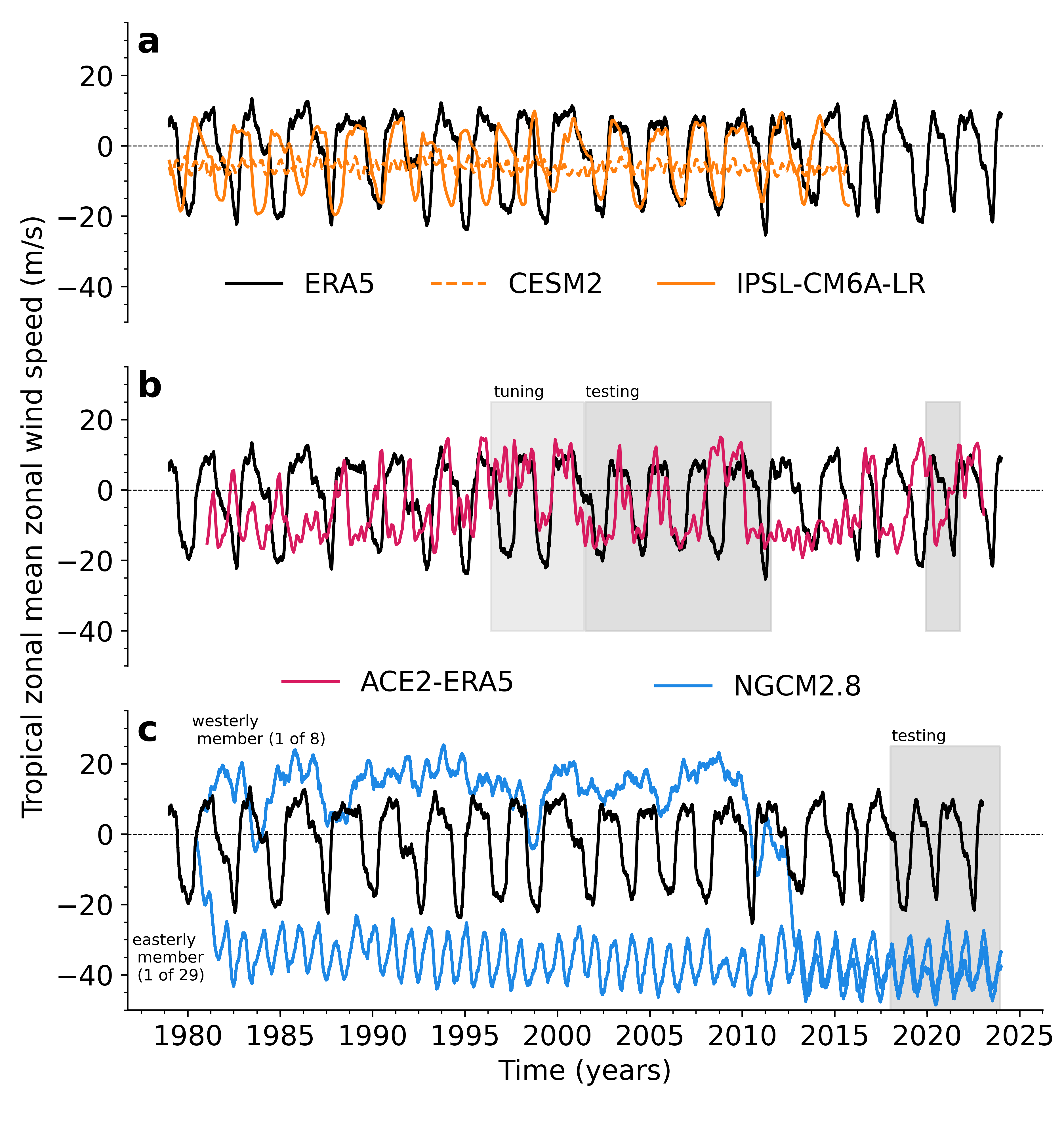}
    \caption{Monthly zonal-mean zonal wind averaged over 10$^{\circ}$N–10$^{\circ}$S from ERA5 (black curves in all panels), (a) two AMIP models (orange curves, solid: IPSL-CM6-LR, dashed: CESM2), (b) one ACE2-ERA5 lagged ensemble member (red curve), and (c) two NeuralGCM2.8 lagged ensemble members (blue curves). Panels a and c show the zonal wind at 50 hPa while panel b shows the average over the vertical layer centered near 50 hPa. The testing periods for each product and the validation period to determine the optimal checkpoint for ACE2-ERA5 are marked in gray shading. See Figure~\ref{fig:Sup. Figure 1} for the peridiocities and amplitudes across each ensemble and Figure~\ref{fig:Sup. Figure 2} for the vertical propagation of zonal winds.}
    \label{fig:Figure 1}
\end{figure}

In ERA5 reanalysis, the QBO period is $\mathrm{\sim}$28 months \cite{pahlavan2021revisiting}, with the winds switching direction from easterly to westerly \cite{baldwinQuasibiennialOscillation2001}; see the black curves in Figure \ref{fig:Figure 1}. These changes in the wind velocity reach up to 40~m/s between the QBO phases (Figure \ref{fig:Figure 1} \& S1b). Physics-based models from AMIP and CMIP6 have been shown to struggle to capture both the periodicity and the amplitude of the QBO \cite{bushellEvaluationQuasiBiennialOscillation2022}. Two models from AMIP (IPSL-CM6A-LR and CESM2), which are with and without QBO-like oscillations, respectively, are shown as representatives of the range across physics-based models in Figure~\ref{fig:Figure 1}a. CESM2 shows a 6-month (semi-annual) oscillation, while IPSL-CM6A-LR shows a 28-month periodicity similar to ERA5. IPSL-CM6A-LR has fine-tuned non-orographic gravity wave drag parameterizations and doubling of vertical levels that help enable it to internally generate QBO signals comparable to ERA5~\cite{boucherPresentationEvaluationIPSLCM6ALR2020}. However, even when AMIP models capture a periodicity close to that in ERA5, they all underestimate the amplitude of the QBO (Figure~\ref{fig:Sup. Figure 1}b), in particular in the lower stratospheric levels below 20 hPa \cite{richterProgressSimulatingQuasiBiennial2020, bushellEvaluationQuasiBiennialOscillation2022}. 

It might be expected that AI emulators and hybrid models are able to overcome some of the issues related to resolution or parameterizations that can lead to biases in QBO statistics in physics-based models. However, the AI emulator and hybrid model examined here are also not able to fully capture the QBO. We show a randomly selected member from the 37-member ACE2-ERA5 lagged ensemble (Figure \ref{fig:Figure 1}b). None of the ensemble members show a regular oscillation in their stratospheric zonal winds, though some members show peaks around 6 months (Figure \ref{fig:Figure 1}b \& S1a). Despite not having a regular periodicity, ACE2-ERA5 does have maximum amplitudes of 39-42 m/s, closer to ERA5 than any of the AMIP models and NeuralGCM (Figure~\ref{fig:Sup. Figure 1}b). ACE2-ERA5 somewhat matches E3SM-1-0 in that they are the only realizations that both overestimate the maximum amplitude and lack a regular periodicity.

NeuralGCM is also unable to produce the timing and amplitude of the QBO. Here, we show stratospheric zonal winds from two members from the 37-member NeuralGCM lagged ensemble (Figure \ref{fig:Figure 1}c). The two members were selected to represent that the NeuralGCM initializations tended to diverge into either an easterly or westerly regime, often dependent on the direction of the winds at initialization. When in the easterly regime, NeuralGCM exhibits a strong annual cycle (Figure \ref{fig:Figure 1}c) superimposed on an excessively strong background flow. Many of the AMIP models that were dominated by 6-month oscillations also remained fixed in an easterly regime. In the westerly regime the model lacks a regular oscillation. In addition, two members out of the 8 that were in the westerly regime abruptly switched to the easterly regime deep into the integration (years 2011, 2015), and changed from irregular fluctuations to an annual cycle when entering the easterly regime. Consistent with the QBO defined by 50 hPa zonal wind (Figure \ref{fig:Figure 1}c), the time-height section of zonal wind in NeuralGCM also does not show downward propagation (Figure~\ref{fig:Sup. Figure 2}c). Rather there are two distinct layers of relatively strong westward zonal winds between 1-5 hPa and 20-50 hPa.

The poor representation of QBO in these two AI models might be due to some of the same issues as in physics-based models—such as limited vertical resolution in the stratosphere, biases in the background flow, or insufficient gravity wave forcing \cite{baldwinQuasibiennialOscillation2001,rindQBOTwoGISS2014}—as well as issues specific to their training process, or a combination of both. Starting with the latter, both AI models are trained to autoregressively predict the next few hours to days via loss function \eqref{eq:loss}, which emphasizes learning the fast dynamics. In emulators such as ACE-ERA5, ``learning'' entirely depends on the definition of the loss function; if errors in the representation of a process do not increase the function, that representation will not be improved. Given the large separation between the QBO's 28-months timescale and loss' 6-hourly timescale, it is likely that slow-varying stratospheric winds do not make a noticeable contribution to the loss function, leading to the poor representation of the main drivers of QBO (i.e., tropospheric gravity waves) and eventually, QBO itself in ACE2-ERA5. This problem can be expected to be less severe for a hybrid model such as NeuralGCM; however, the same time scale separation can lead to a lack of learning gravity waves in the model's neural network-based parameterization. Furthermore, AI models often use a bottom-heavy weighting of the loss functions for certain variables to improve the representation of near-surface variables (e.g., humidity), which can further lead to poor learning of the stratosphere. 

As noted above, the problem may also be, to some extent, due to the lack of vertical resolution in the stratosphere. ACE2-ERA5 has only one vertically averaged level with its lower interface at 50 hPa representing the stratosphere \cite{watt-meyerACE2AccuratelyLearning2025a}. NeuralGCM has finer vertical resolution (11 of 37 levels above 100 hPa), but its loss function is applied only up to 30 hPa, leaving just 4 stratospheric levels directly constrained. The vertical structure of zonal winds does not show any improvement at levels below 30 hPa that were included in training compared with the upper stratosphere (Figure~\ref{fig:Sup. Figure 2}c). Additional training tests that account for more vertical levels in the stratosphere are needed to fully evaluate the capabilities of AI emulators and hybrid models in the stratosphere.  

\subsection{Convectively coupled equatorial waves}
\begin{figure}
\noindent\includegraphics[width=\textwidth]{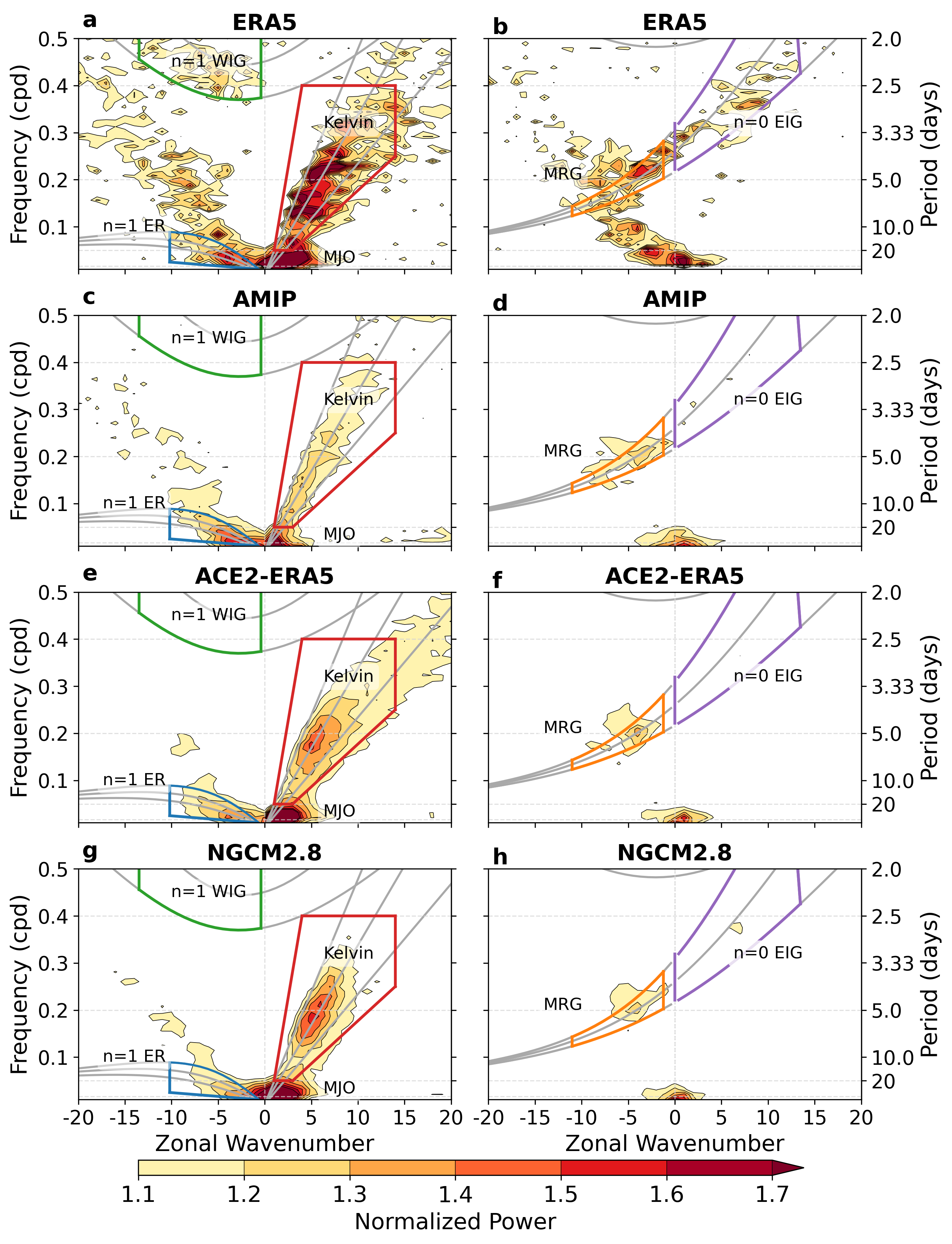}
\caption{Wavenumber-frequency power spectrum of the symmetric (left column) and antisymmetric (right column) components of daily mean precipitation from (a-b) ERA5, (c-d) one AMIP model (CESM2-WACCM), (e-f) ACE2-ERA5, and (g-h) NeuralGCM. See Figure~\ref{fig:Sup. Figure 3} for the background spectra.}
\label{fig:Figure 2}
\end{figure}

We next evaluate tropical atmospheric variability associated with convectively coupled equatorial waves (i.e., MJO, Rossby, Kelvin, MRG, and IG  waves) by computing wavenumber-frequency spectra of tropical precipitation or total column moisture convergence (Figure~\ref{fig:Figure 2}).  

CESM2-WACCM (CAM6) is used as a representative AMIP simulation in our examination of convectively coupled waves. This physics-based model has been shown to generate eastward propagation of convectively coupled precipitation and wind patterns \cite{danabasogluCommunityEarthSystem2020} and here shows skill in capturing the key signals in the MJO, Rossby wave, and Kelvin wave regions of the spectra (Figure \ref{fig:Figure 2}c-d). However, the signals associated with MJO between wavenumbers 2 to 5 and Kelvin waves are relatively weak compared with ERA5 (Figure \ref{fig:Figure 2}c). CESM2-WACCM also shows very little spectral power in the MRG and IG regions (Figure \ref{fig:Figure 2}d).

ACE2-ERA5 and NeuralGCM also show an overall good agreement in their tropical wave-frequency spectra compared to ERA5 (Figure~\ref{fig:Figure 3}e-h). Both are able to generate equatorial Rossby waves, Kelvin Waves, and the MJO signals, though they underestimate the power in the higher frequency Kelvin wave and higher wavenumber Rossby wave regions of the spectra. Compared with CESM2-WACCM, the AI emulator and hybrid model show an improvement in the representation of the MJO between wavenumbers 2 and 5. AMIP, ACE2-ERA5, and NeuralGCM all show some power in the westward propagating MRG wave regions, but underestimate the asymmetric spectra overall, particularly the higher-frequency eastward-propagating IG waves (Figure \ref{fig:Figure 2}, right column). In general, both the AMIP and AI models underestimate the signals at higher frequencies. 

The fast timescale processes on the order of 6-hourly to daily timesteps may be most important for capturing the propagation of equatorial convectively coupled waves. This version of NeuralGCM, without full prognostic precipitation, underestimates the power in the higher frequency Kelvin wave region of the spectra similar to CESM2-WACCM, whereas ACE2-ERA5, which prognostically forecasts precipitation, is able to capture more of the power in this region.

Several studies have hypothesized that tropical convection, particularly the MJO, is modulated by the phase of the QBO \cite{baldwinQuasibiennialOscillation2001,sonPreferredModesVariability2006,sonStratosphericControlMadden2017,zhangQBOMJOConnection2018}. Yet, many physics-based models that generate both the QBO and MJO internally do not consistently capture their teleconnection. Here, although neither AI model produces a QBO, we conducted a preliminary analysis of the MJO conditioned on easterly versus westerly equatorial zonal winds at 50 hPa. We found no significant differences (not shown). Thus, whether and how the QBO and MJO interact in future AI emulators and hybrid models that better capture stratospheric variability remains an open question.

\subsection{Extratropical wave-mean flow interaction}

 Extratropical variability is dominated by eddy-mean flow interactions. These interactions involve eddies being limited in latitude by the critical line (where the mean flow equals the eddy phase velocity) in a highly nonlinear wave-breaking process. We quantify the extratropical eddy-mean flow interactions by computing Southern Hemisphere momentum fluxes during DJFM and JJAS as a function of angular phase speed and latitude (Figure 3) \cite{hayashiGeneralizedMethodResolving1971,randelPhaseSpeedSpectra1991,chenPhaseSpeedSpectra2007}. In ERA5, during both seasons, equatorward propagating baroclinic waves break along a narrow region bounded by the critical line or latitude (purple lines in Figure 3a,e).

\begin{figure}
\noindent\includegraphics[width=\textwidth]{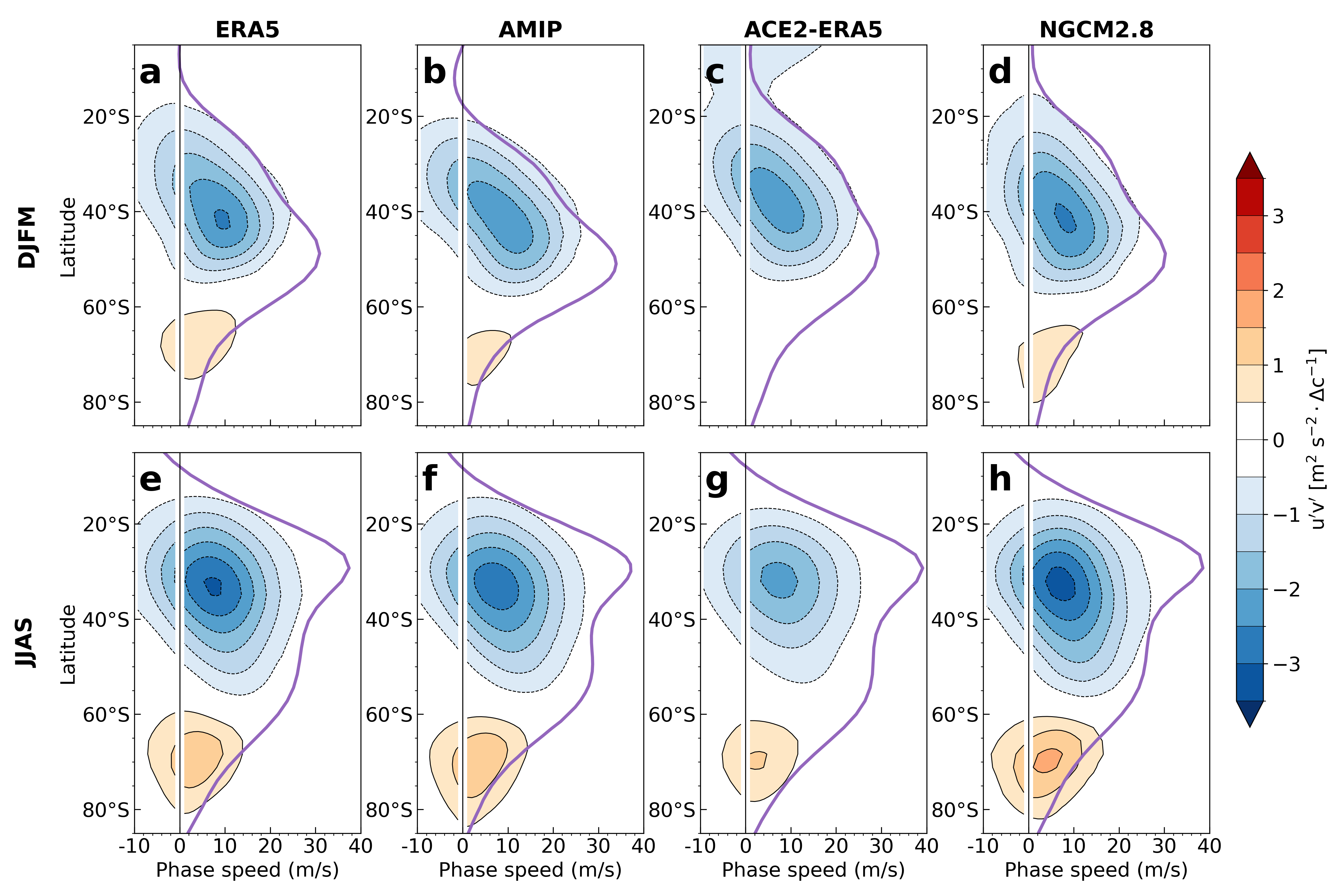}
\caption{Contours of 250 hPa transient eddy momentum flux versus latitude and phase speed for DJFM (left column) and JJAS (right column) from (a-b) ERA5, (c-d) one AMIP model (CESM2-WACCM), (e-f) ACE2-ERA5 (37 member mean), and (g-h) NeuralGCM2.8 (37 member mean). Shading intervals are 0.50 $\mathrm{m^{2}\ s^{-2} \cdot \Delta c^{-1}}$. The eddy momentum fluxes are normalized by phase speed bin size ($\Delta c$). Purple contours in each panel denote seasonally averaged mean zonal wind. See Figure~\ref{fig:Sup. Figure 4} for eddy momentum flux convergence.}
\label{fig:Figure 3}
\end{figure}

The eddies primarily propagate eastward at angular phase speeds between 5-20 m/s, and weakly eastward poleward of 60 $\mathrm{^{\circ}S}$, with the most prominent signal near the center of the midlatitude jet at around 10 m/s (Figure \ref{fig:Figure 3}). All patterns show the peak of eddy momentum flux and convergence in midlatitudes near the center of the jet, with equatorward propagation south of 55$\mathrm{^{\circ}S}$ and poleward propagation north of 55 $\mathrm{^{\circ}S}$ (Figure~\ref{fig:Sup. Figure 4}). Linear wave theory suggests that where the zonal mean wind equals the phase speed $\overline{u}=c$, again referred to as the critical latitude, the baroclinic waves should break or dissipate, accelerating the jet. Both AMIP models and the AI models are able to capture these complex features, such as the eddy flux (shading) being bounded to angular phase speeds less than the mean zonal wind (purple line) (Figure \ref{fig:Figure 3}). This is most clearly seen in the austral summer (top row in Figure \ref{fig:Figure 3}), with all models able to capture where the eddy flux contours are parallel to the mean zonal wind or critical line.

However, the representative AMIP model (CESM2-WACCM) and ACE2-ERA5 underestimate the spectral density of eddy momentum flux seen in the reanalysis, despite similar mean wind fields (Figure \ref{fig:Figure 3}). In contrast, NeuralGCM more closely matches the pattern and slightly overestimates the amplitude relative to ERA5, suggesting that the hybrid approach may yield some improvements in representing the strength of eddy momentum fluxes (this advantage is also visible in Fig. S4). As we compute the eddy flux cospectra across each year individually, we compare the testing and training periods for each AI model and do not find noticeable differences. 

\subsection{Propagation of the Southern Annular Mode (SAM)} 

\begin{figure}
    \centering
    \noindent\includegraphics[width=\textwidth]{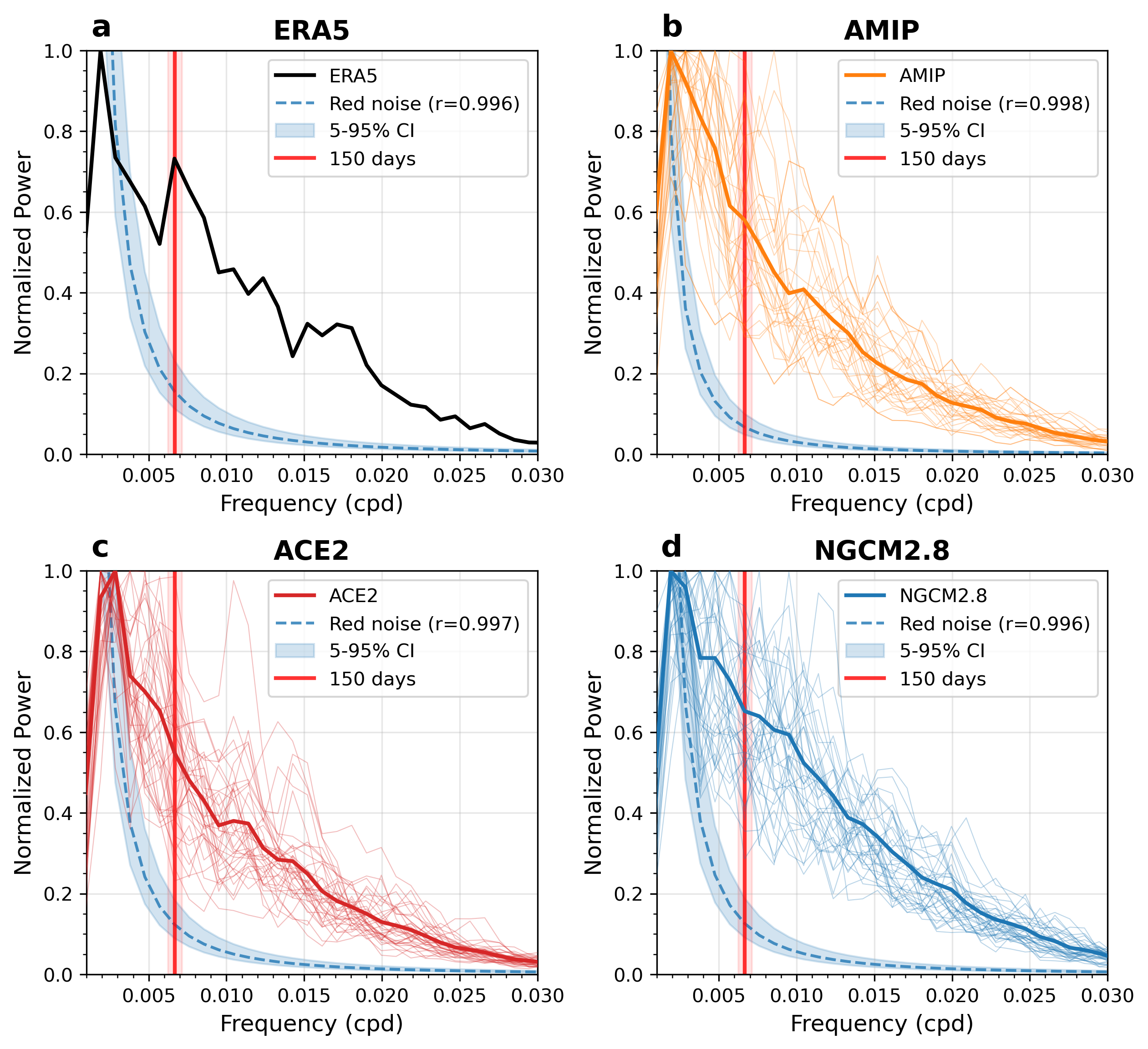}
    \caption{Frequency (cycles per day) power spectra for 80$\mathrm{^{\circ{}}}$S to 20$\mathrm{^{\circ{}}}$S zonal mean zonal wind anomalies projected onto the first leading EOF mode ($z_{1}$) from (a) ERA5, (b) AMIP, (c) ACE2-ERA5, and (d) NeuralGCM. Thick lines represent ensemble means and thin lines represent individual models or realizations. The dashed blue line and shaded region denote the red noise curve and its 95\% confidence interval. The vertical red line denotes the 150-day periodicity associated with the SAM.}
    \label{fig:Figure 4}
\end{figure}

Our final benchmark evaluates skill in the spatial pattern and timing of the poleward propagation of SAM, which is the dominant mechanism of variability determining Southern Hemisphere climate and generally defined as the leading EOF mode of sea level pressure or winds \cite{kidsonIndicesSouthernHemisphere1988,sonPreferredModesVariability2006, lubisIntrinsic150DayPeriodicity2023}. Sheshadri et al. (2017)~\cite{sheshadriPropagatingAnnularModes2017} showed that the first two leading modes ($z_{1}$ and $z_{2}$) could be considered two phases of a single dynamical system representing the poleward migration of zonal mean zonal wind anomalies ($z(t)$ is the timeseries of the EOF mode). 
From this framework, Lubis et al. (2021)~\cite{lubisEddyZonalFlow2021} and Lubis and Hassanzadeh (2023)~\cite{lubisIntrinsic150DayPeriodicity2023} later highlighted the importance of eddy-zonal flow interactions for describing the SAM dynamics and its influence on Southern Hemisphere climate; they also predicted and verified a 150-day periodicity in SAM variability (Figure \ref{fig:Figure 4}a). Fully-coupled CMIP6 models were found to be overly persistent, underestimating the cross-EOF correlations that represent the strength of eddy-zonal flow feedbacks, and mischaracterizing the periodicity~\cite{lubisIntrinsic150DayPeriodicity2023} (purple lines in Figure~\ref{fig:Sup. Figure 5}a-b).

We find that AMIP and the AI emulator and hybrid model are able to capture the cross EOF correlations or eddy feedbacks, as accurately or better than in the CMIP6 ensemble mean (Figure~\ref{fig:Sup. Figure 5}c-f). Consistent with the results from the eddy momentum fluxes (Section 3.3), ACE2-ERA5 shows comparable over-persistence of the leading mode compared with physics-based models (Figure~\ref{fig:Sup. Figure 5}a-b vs. Figure~\ref{fig:Sup. Figure 5}c-f), but slightly overestimates the lagged correlations in the transition from $\mathrm{z_{2}}$ to $\mathrm{z_{1}}$ while underestimating the transition in the opposite direction. This can also be seen in the latitudinal structure of the poleward propagating wind field, where the correlations with the base latitude diverge at longer lag times (Figure~\ref{fig:Sup. Figure 6}). NeuralGCM very accurately captures the autocorrelation of the first leading mode and the transition from $\mathrm{z_{2}}$ to $\mathrm{z_{1}}$ (Figure~\ref{fig:Sup. Figure 5}). 

Though the AMIP models and AI emulator and hybrid model capture comparable cross EOF feedbacks as ERA5, they do not display a regular or coherent spectral peak at 150 days associated with the progapation of SAM across models or ensemble members (Figure \ref{fig:Figure 4}). Similar to the QBO spectra (Figure~\ref{fig:Sup. Figure 1}b), there is not a consistent spectral peak across ensemble members of the AI emulator and hybrid model that coincides with the 150 periodicity derived from theory or the well-defined peak at 150 days shown in ERA5 (Figure \ref{fig:Figure 4}a). When averaging across either the AMIP or AI emulator or model ensembles, there is no clear spectral peaks. Despite most accurately representing the $\mathrm{z_{1}}$ autocorrelation and $\mathrm{z_{2}}$ to $\mathrm{z_{1}}$ transition, NeuralGCM still does not consistently exhibit the 150 day peak across its ensemble members. (Figure \ref{fig:Figure 4}d). 

Overpersistence in physics-based models is often attributed to the representation of resolution or parameterizations of diabatic processes~\cite{scaifeSignaltonoiseParadoxClimate2018, smithDiabaticEddyForcing2024,weisheimerSignaltoNoiseParadoxClimate2024}. AI emulators and hybrid models could theoretically overcome these limitations through the use of neural networks; however they exhibit many of the same discrepancies as physics-based models, such as overpersistence and weak eddy-mean flow feedbacks. For example, they show an ability to capture the climatological patterns of extratropical eddy momentum flux (Figure \ref{fig:Figure 2}) that control the shifts between phases of the SAM. The AI emulator and hybrid model are able to capture aspects of complex interactions between eddies and the mean flow as well or better than the physics-based models, but potential limitations remain, such as time scale separation and weighting or loss functions in the training, that prevent the AI emulator and hybrid model from consistently capturing the 150-day oscillation of the SAM~\cite{bonavitaLimitationsCurrentMachine2024, kentSkilfulGlobalSeasonal2025}. 

\section{Conclusions \& Discussion}
    
\subsection{Conclusions}

We evaluate the performance of an AI emulator (ACE2-ERA5) and hybrid model (NeuralGCM), both trained on ERA5 reanalysis and forced with observed sea surface temperatures and sea ice concentrations, relative to physics-based AMIP simulations and ERA5. The set of metrics used here can be considered strict tests of complex atmospheric dynamics, including eddy-mean flow interactions. Overall, the AI atmospheric hybrid model and emulator capture short time scale tropospheric variability, such as convectively coupled equatorial waves represented by tropical precipitation and extratropical wave-mean flow interaction represented by wind spectra, over the satellite era. However, they struggle in two major aspects related to longer timescale variability, such as the generation of QBO-like fluctuations in the stratosphere and the propagation of the SAM, potentially related to the AI models' loss function focused on fast (weather-scale) dynamics.

Neither the dynamical models nor the AI emulators reproduce the full characteristics of a QBO-like oscillation in their 50~hPa zonal winds. This has been a persisting challenge in physics-based climate models \cite{orbeRepresentationModesVariability2020}, and it appears that the AI models are also subjected to similar limitations. The causes of this deficiency may be related to lack of learning of gravity waves due to the design of the loss function and its weighting, and/or lack of  vertical resolution in the stratosphere. Additional analyses and tests of other training approaches (e.g., offline-online learning~\cite{pahlavan2024explainable}) is necessary to determine the cause(s) in NeuralGCM and ACE2-ERA5.

\subsection{Discussion}

NeuralGCM and ACE2-ERA5 evaluated here show high skill in capturing complex higher-frequency and slower (out to 60-days) atmospheric variability, including tropical waves and extratropical eddy-mean flow interactions, but struggle with longer timescale processes (QBO, SAM). This adds to the growing line of evidence that AI models can learn some degree of dynamics rather than simply memorizing patterns; however, they currently have major shortcomings for tasks beyond short- and medium-term forecasting   \cite{hakimDynamicalTestsDeep2024, bano-medinaAreAIWeather2025, sunCanAIWeather2025, vanloonReanalysisBasedGlobalRadiative2025}. In particular, it has been suggested that accurate representation of the atmospheric modes of variability (such as those examined in this study) is important for accurate projections of the future response to anthropogenic forcing~\cite{gerberAnnularModeTime2008, shepherdAtmosphericCirculationSource2014}. It is possible that the inability to capture these longer timescale processes may limit the AI models' ability to accurately project the response to anthropogenic forcing or represent yet unseen climates -- an important success of physics-based models~\cite{manabeSensitivityGlobalClimate1980}. 

In all, these shortcomings of AI emulator and hybrid model may be addressed to some extent with updates to the training procedure and architecture of the models. Note that biases, especially in processes such as QBO, have persisted across generations of physics-based AMIP models \cite{richterProgressSimulatingQuasiBiennial2020}. AI models have a large number of parameters (the neural networks' weights) that can be tuned/calibrated to reduce biases in targeted metrics, e.g., using reinforcement learning or ensemble Kalman inversion \cite{laiMachineLearningClimate2025,pahlavan2024explainable}. However, this even further enhances the importance of in-depth dynamical tests to ensure that the models are doing the right things for the right reasons.

Here, we focused on the only two available AI models that are trained on ERA5 with an AMIP configuration. Given the rapid emergence of various AI models, especially emulators, we expect more models available for testing in the future \cite{dheeshjithSamudraAIGlobal2025, chapmanCAMulatorFastEmulation2025}. As for the two current models, the results shown here could be associated with specific model choices made during training (e.g., different seeds, checkpoints, weighting, optimization for global quantities)~\cite{kochkovNeuralGeneralCirculation2024, watt-meyerACE2AccuratelyLearning2025a}. Going forward, the skill and fidelity of atmospheric hybrid models and emulators to be used for climate studies can be strongly determined by these choices in training, such as better consideration of the stratosphere through vertical weighting or resolution. Also, and perhaps most importantly, it is likely that optimizing on weather timescales, i.e., short rollouts (hours to days), may come at the expense of longer timescale processes necessary to make climate projections. Continued evaluation such as that shown here will be necessary to determine the best approaches for training and model architecture.

A key takeaway that might inform the direction of future model development is the comparison between the fully data-driven and hybrid modeling approaches. Overall, we do not find significant differences between ACE2-ERA5 and NeuralGCM in their ability to represent atmospheric modes of variability over the historical period. The physical constraints imposed by the dynamical core might have been expected to better enable the model to capture important dynamics for simulating these modes of variability, especially at longer timescales. However, the two AI models exhibit very similar results, suggesting that data-driven methods might be capable of learning the key interactions simulated by the dynamical core. Whether or not this applies to other hybrid models, and more importantly, to out-of-distribution forcing, remains an open question. At the moment there are only two available models that provide the necessary outputs for this analysis and the introduction of more AI atmosphere and climate models, such as DLESyM \cite{cresswell-clayDeepLearningEarth2025}, will be key in addressing confirming the robustness of these conclusions. 

There have been many other successes and lessons learned from decades of development of physics-based climate models beyond what was included here that should be applied to the rapidly developing field of AI climate emulators~\cite{balajiClimbingCharneysLadder2021a}. In particular, the use of a hierarchy of test systems has been essential to improve models and understanding, and can also play a key role here \cite{jeevanjee2017perspective,bordoni2025futures,ullrich2025recommendations}. The benchmarking metrics and evaluation shown in this study provide just an initial assessment of the performance of AI atmospheric models, an essential step towards using these models for any kind of out-of-distribution application (i.e., for unseen future climates). These four benchmarking metrics were chosen based on the available output from existing AI emulators or hybrid models. Future evaluation will be greatly benefited by a broader representation of variables and processes, in addition to greater (vertical) resolution. 

Though the use of AI emulators and hybrid models for global atmosphere emulation is still in its early stages, our evaluation highlights the immense progress that has been made in AI-based modeling in a very short time, leading to models that can already capture many complex aspects of atmospheric dynamics fairly well, most strikingly, just from data.

%
%

\section*{Open Research Section}
ERA5 is publicly available from the Copernicus Climate Change Service (C3S) Climate Data Store (CDS) at \url{https://cds.climate.copernicus.eu/cdsapp#!/dataset/}. CMIP6 AMIP data are available from the Earth System Grid Federation (ESGF) at \url{https://esgf-node.llnl.gov/projects/cmip6/}. ACE2-ERA5 is available from the Allen Institute for AI at \url{https://huggingface.co/allenai/ACE2-ERA5}. NeuralGCM is available from \url{https://github.com/neuralgcm/neuralgcm}.
Code and data to remake figures and compute diagnostics will be made available upon publication.

\section*{Acknowledgments}
IB and TAS are supported by the National Oceanic and Atmospheric Administration award NA23OAR4310597. HP, PH, and KR are supported by National Science Foundation grant OAC-2544065 and by Schmidt Sciences, LLC. Computational resources by NSF ACCESS (ATM170020), NCAR’s CISL (URIC0009), and the University of Chicago Research Computing Center. 

%
\bibliographystyle{unsrt}  
\bibliography{Circulation_Variability}


\clearpage
\appendix
\section*{Supplementary Information (SI)}

\renewcommand{\thefigure}{S\arabic{figure}} 
\setcounter{figure}{0}

\begin{table}
    \centering
    \caption{List of AMIP models used in this study.}
    \begin{tabular}{ c c c c}
        \hline
        \multicolumn{4}{ c }{AMIP models} \\
        \hline
        Model Name & Institution & Realizations & Variables (daily mean) \\
        \hline
        ACCESS-CM2 & CSIRO & r1i1p1f1 & zonal wind (ua), \\
        & & & meridional wind (va) \\
        ACCESS-ESM1-5 & CSIRO & r1i1p1f1 & ua, va \\
        BCC-CSM2-MR & BCC & r2i1p1f1, r3i1p1f1 & ua \\
        CAMS-CSM1-0 & CAMS & r1i1p1f1, r2i1p1f1, r3i1p1f1 & ua, va \\   
        CESM2-FV2 & NCAR & r1i1p1f1 & ua, va \\
        CESM2-WACCM & NCAR & r1i1p1f1 & ua, va, \\
        & & & precipitation rate (pr) \\
        CESM2-WACCM-FV2 & NCAR & r1i1p1f1, r2i1p1f1, r3i1p1f1 & ua, va \\
        CESM2 & NCAR & r1i1p1f1, r10i1p1f1 & ua \\
        CIESM & & r1i1p1f1 & ua \\
        CMCC-CM2-HR4 & CMCC & r1i1p1f1 & ua, va\\
        CMCC-CM2-SR5 & CMCC & r1i1p1f1 & ua, va \\
        CNRM-ESM2-1 & CNRM & r1i1p1f1 & ua, va \\
        CNRM-CM6-1-HR & CNRM & r1i1p1f1 & ua \\
        CNRM-CM6-1 & CNRM & r1i1p1f1 & ua \\
        CanESM5 & CCCma & r1i1p1f1, r1i1p2f1, r2i1p1f1, & ua, va \\
        & & r2i1p2f1, r3i1p2f1, & \\
        & & r4i1p2f1, r5i1p2f1 & \\
        E3SM-1-0 & DOE & r1i1p1f1, r2i1p1f1 & ua \\
        EC-Earth3 & EC-Earth & r1i1p1f1 & ua, va\\
        EC-Earth3-CC & EC-Earth & r1i1p1f1 & ua, va \\
        FGOALS-f3-L & CAS & r1i1p1f1, r2i1p1f1, r3i1p1f1 & ua, va \\
        FGOALS-g3 & CAS & r1i1p1f1 & ua, va \\
        GFDL-CM4 & NOAA-GFDL & r1i1p1f1 & ua, va \\
        GFDL-AM4 & NOAA-GFDL & r1i1p1f1 & ua \\
        GFDL-ESM4 & NOAA-GFDL & r1i1p1f1 & ua \\
        HadGEM3-GC31-LL & MOHC & r1i1p1f3, r2i1p1f3, r3i1p1f3, & ua, va \\
        & & r4i1p1f3, r5i1p1f3 & \\
        HadGEM3-GC31-MM & MOHC & r1i1p1f3, r2i1p1f3 & ua \\
        IITM-ESM & IITM & r1i1p1f1 & ua, va \\
        INM-CM4-8 & INM & r1i1p1f1 & ua, va \\
        INM-CM5-0 & INM & r1i1p1f1 & ua, va \\
        IPSL-CM6A-LR & IPSL & r1i1p1f1, r2i1p1f1, r3i1p1f1, & ua \\
        & & r4i1p1f1, r51p1f1, r6i1p1f1, & \\
        & & r7i1p1f1, r8i1p1f1, r9i1p1f1, & \\
        & & r10i1p1f1 & \\
        KACE-1-0-G & & r1i1p1f1 & ua \\
        MIROC6 & MIROC & r1i1p1f1, r2i1p1f1, r3ip1f1, & ua \\
        & & r3ip1f1, r4i1p1f1, r5i1p1f1, & \\
        & & r6i1p1f1, r7i1p1f1, r8i1p1f1, & \\
        & & r9i1p1f1 & \\
        MIROC-ES2L & MIROC & r1i1p1f1 & ua, va \\
        MPI-ESM-1-2-HAM & MPI-M & r1i1p1f1 & ua, va \\
        MPI-ESM1-2-HR & MPI-M & r1i1p1f1 & ua, va \\
        MPI-ESM1-2-LR & MPI-M & r1i1p1f1 & ua, va \\
        MPI-ESM2-0 & MPI-M & r1i1p1f1, r2i1p1f1, r3i1p1f1 & ua \\
        NESM3 & NUIST & r1i1p1f1, r2i1p1f1, r3i1p1f1 & ua, va \\
        & & r4i1p1f1, r5i1p1f1 & \\
        NorCPM1 & NCC & r1i1p1f1 & ua \\
        NorESM2-LM & NCC & r1i1p1f1 & ua, va \\
        SAM0-UNICON & SNU & r1i1p1f1 & ua, va \\  
        TaiESM1 & & r1i1p1f1 & ua \\
        UKESM1-0-LL & MOHC & r1i1p1f1 & ua \\ \\
        \hline
    \end{tabular}
    \label{table: Models list}
\end{table}

\clearpage

\begin{figure}
    \centering
    \includegraphics[width=\textwidth]{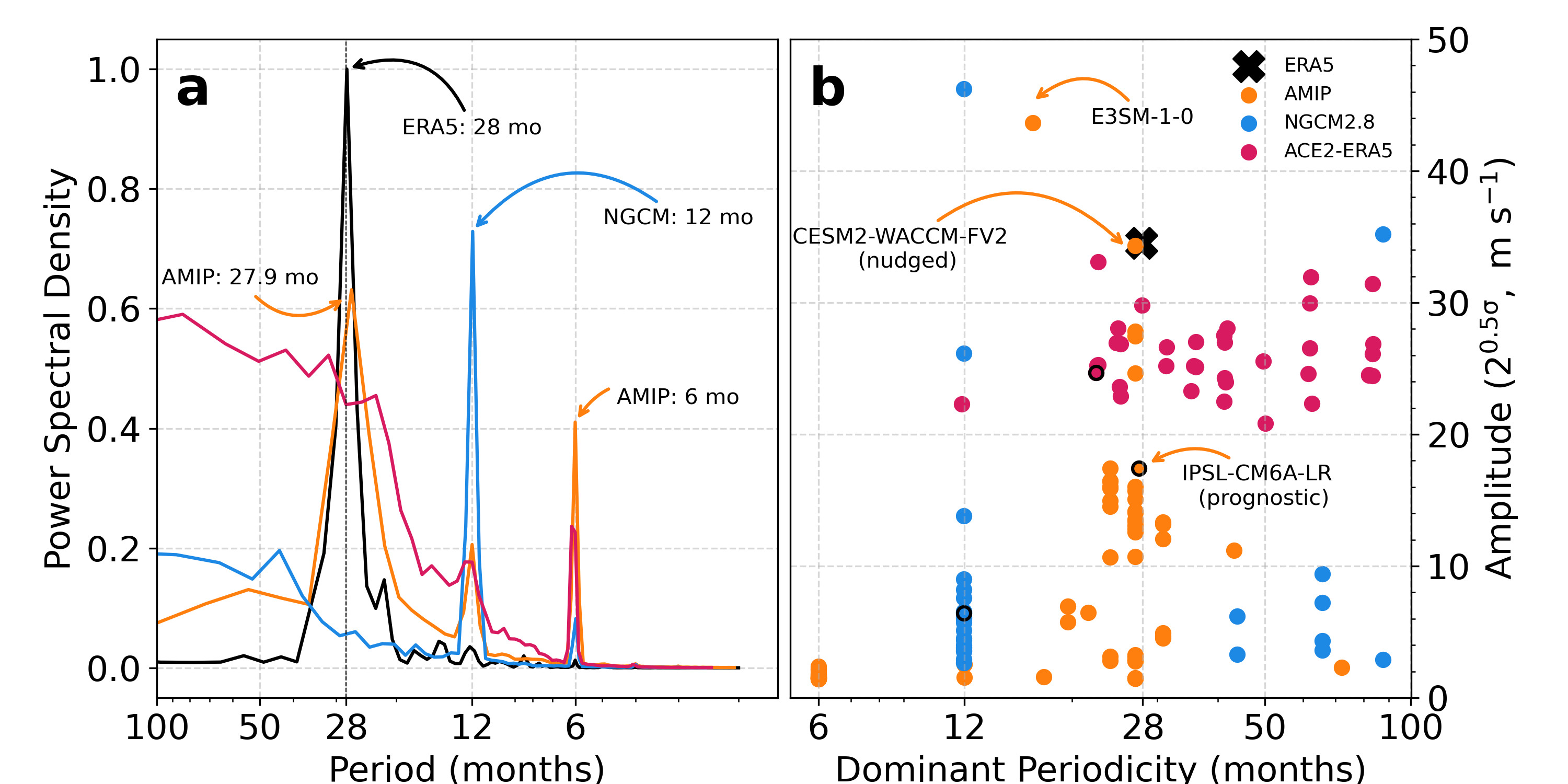}
    \caption{(a) Power spectra of 50-hPa zonal-mean zonal wind in the tropics ($\mathrm{10^{\circ}S-10^{\circ}N}$) from ERA5 (black), AMIP (orange), NeuralGCM2.8 (blue), and ACE2-ERA5 (red). The power spectra for each ensemble member is computed individually, and the ensemble average is shown in panel (a). (b) Maximum zonal-mean zonal wind amplitude (maximum minus minimum zonal mean wind) versus each model/realization's dominant period.}
    \label{fig:Sup. Figure 1}
\end{figure}

\begin{figure}
    \centering
    \includegraphics[width=\textwidth]{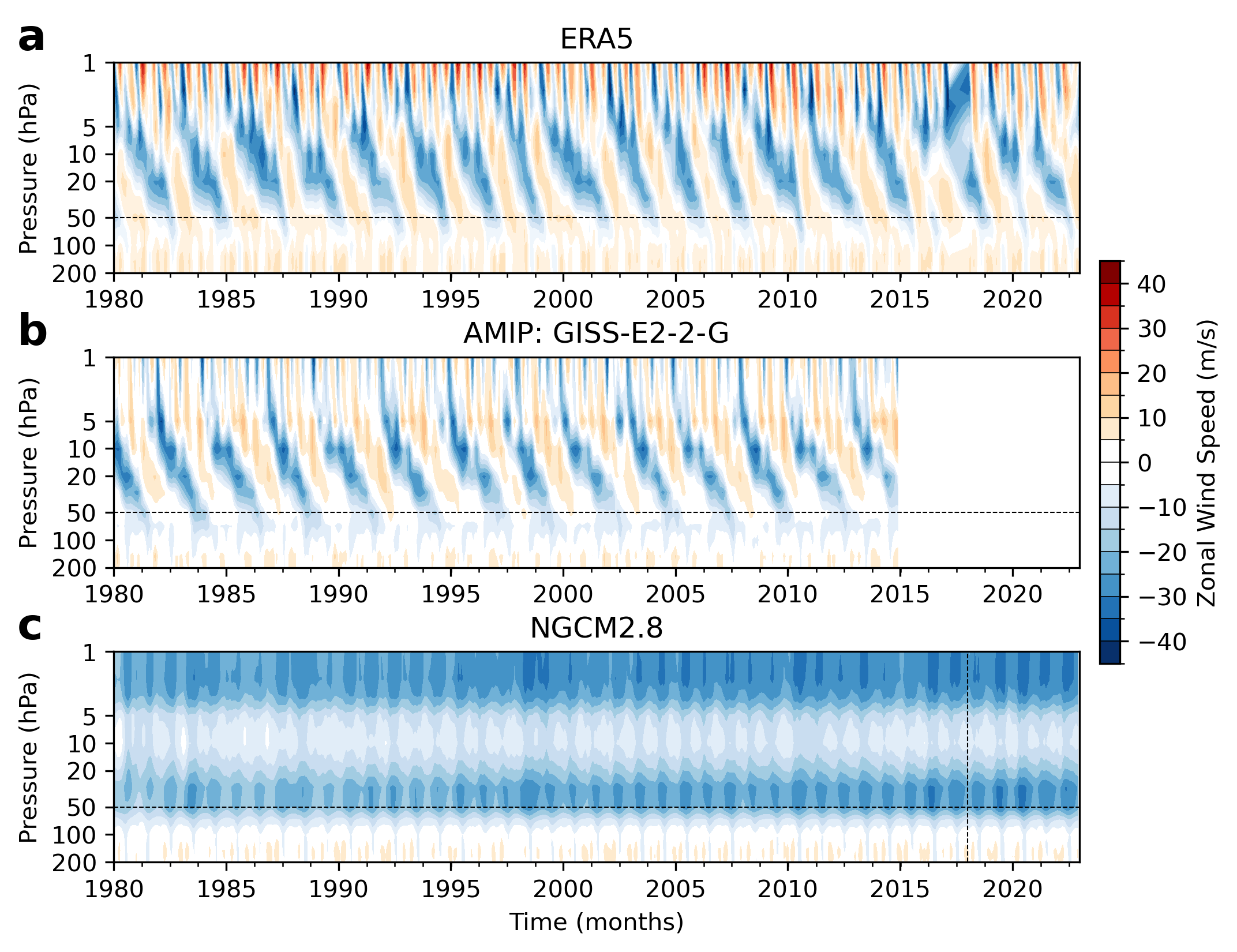}
    \caption{Hovmoller diagrams of equatorial ($\mathrm{10^{\circ}S - 10^{\circ}N}$) mean zonal winds from (a) ERA5, (b) GISS-E2-2G, and (c) NeuralGCM2.8. The vertical dashed black line in panel (c) delineates the training and testing periods.}
    \label{fig:Sup. Figure 2}
\end{figure}

\begin{figure}
    \centering
    \includegraphics[width=\textwidth]{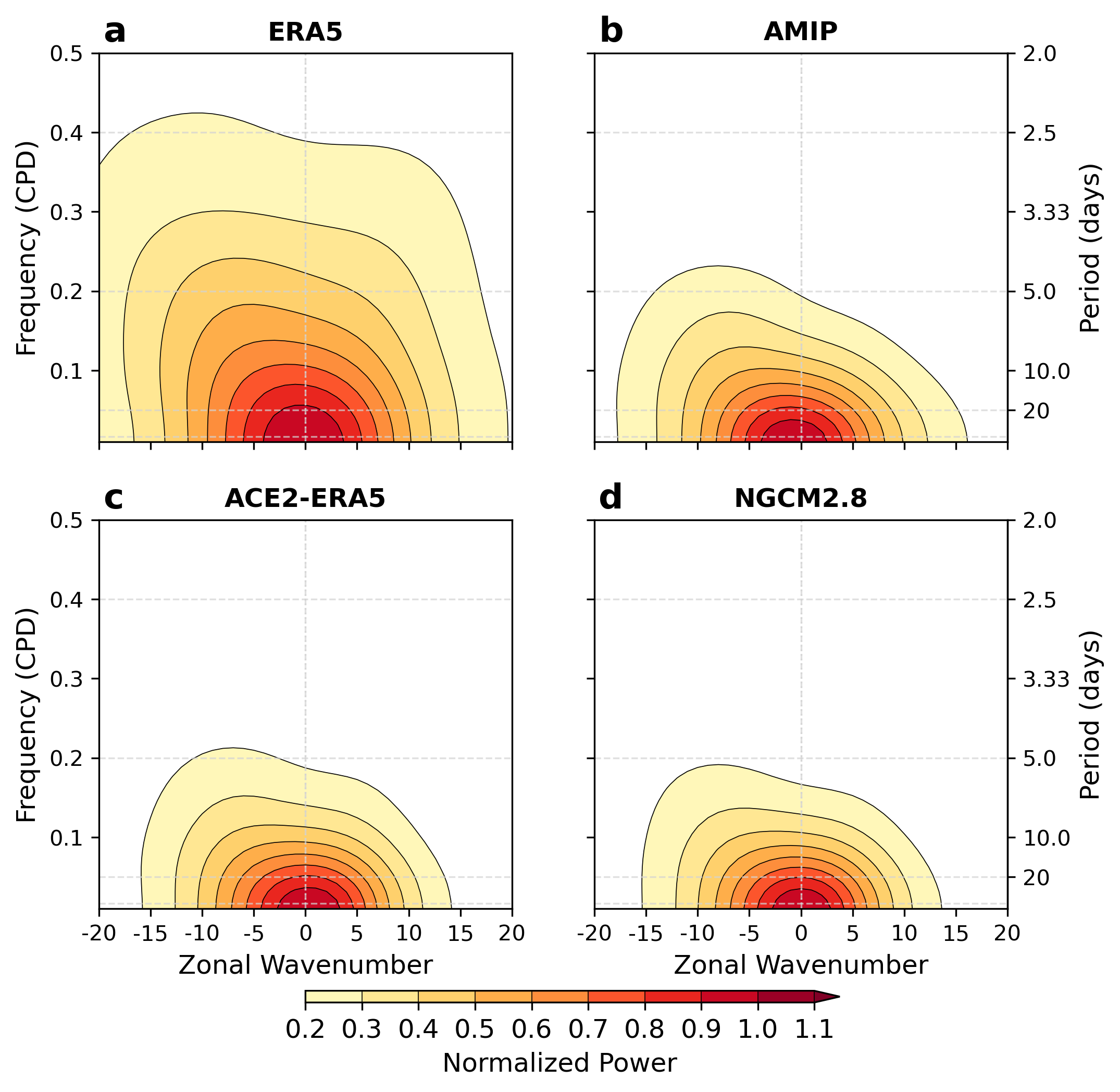}
    \caption{Same as Fig. 4 but showing the background spectra.}
    \label{fig:Sup. Figure 3}
\end{figure}

\begin{figure}[p]
    \centering
    \includegraphics[width=\textwidth]{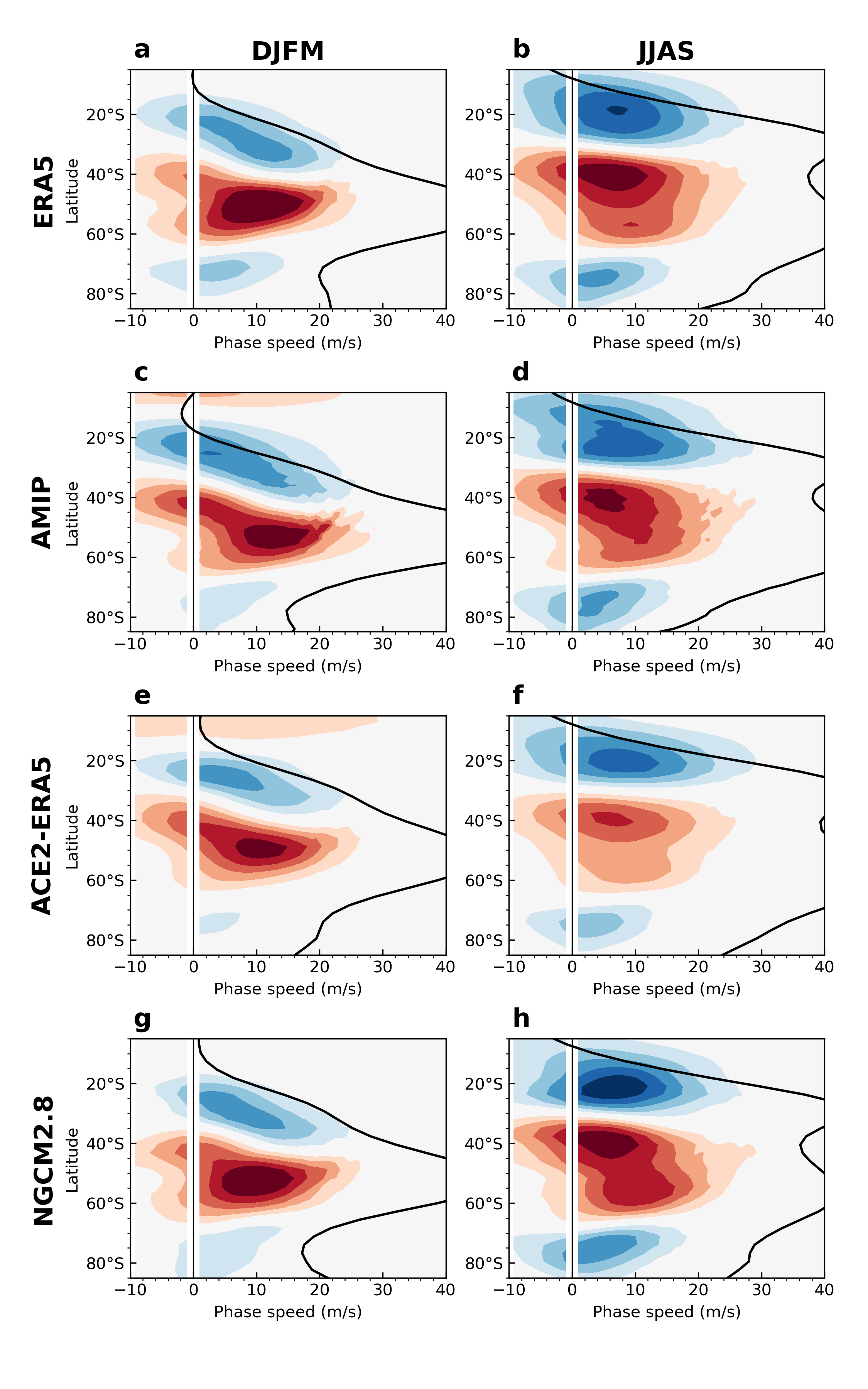}
    \caption{\small
    (a) Same as Fig. 3 but using eddy momentum flux convergence, following Chen and Held (2007) rather than $uv$ transient power as in Randel and Held (1991)~\cite{randelPhaseSpeedSpectra1991}. Shading is shown on intervals of 0.02 m/s/day.}
    \label{fig:Sup. Figure 4}
\end{figure}

\clearpage

\noindent{Figure S4: (a) Same as Fig. 3 but using eddy momentum flux convergence, following Chen and Held (2007) rather than $uv$ transient power as in Randel and Held (1991)~\cite{randelPhaseSpeedSpectra1991}. Shading is shown on intervals of 0.02 m/s/day.} 
\clearpage

\begin{figure}
    \includegraphics[width=\textwidth]{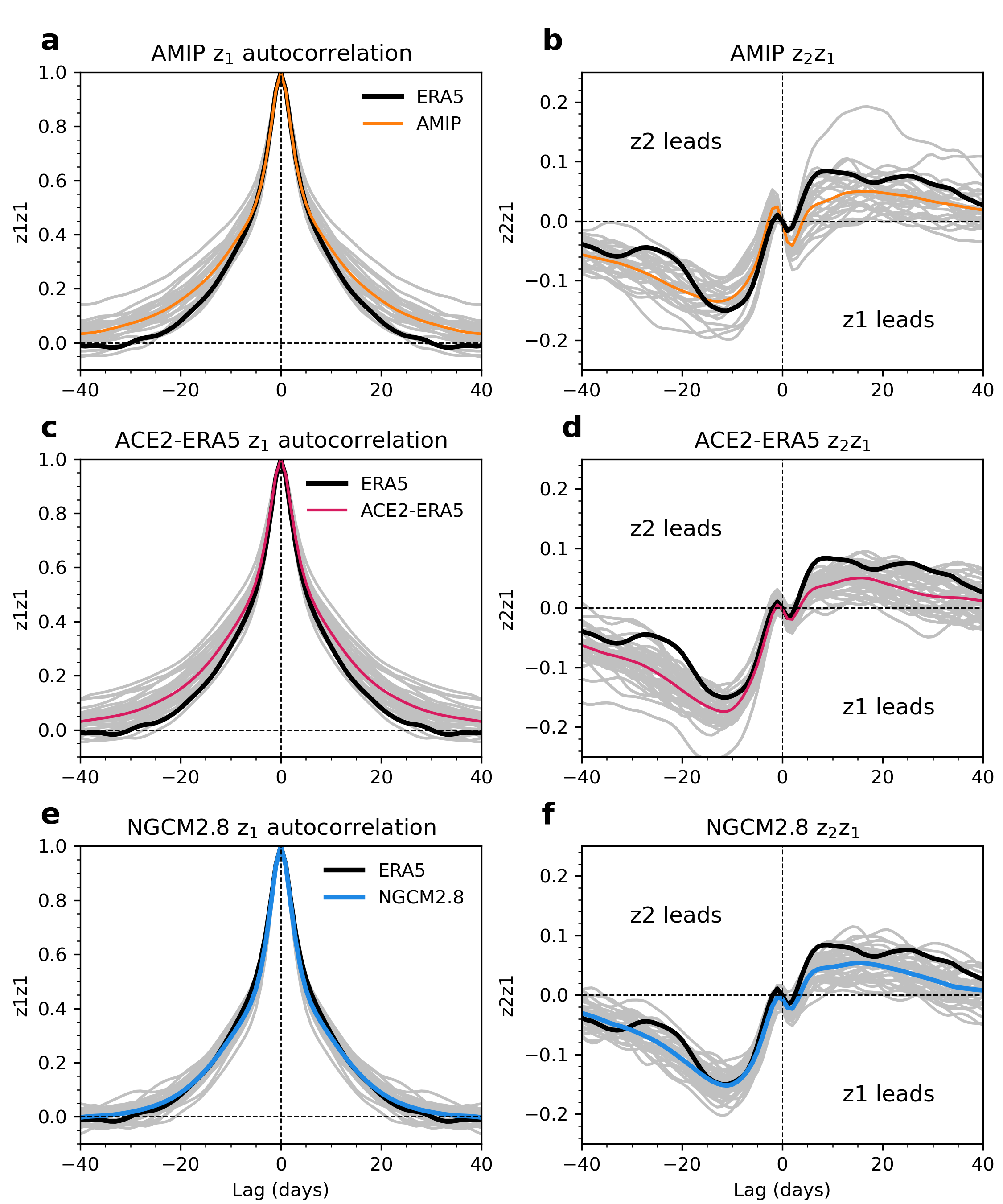}
    \caption{Autocorrelation of the leading EOF mode ($\mathrm{z_{1}z_{1}}$, left column) and the lagged correlation between the second and first leading mode ($\mathrm{z_{2}z_{1}}$, right column) of anomalous daily averaged mean 500 hPa zonal wind from 80--20 $^{\circ}$S using (a-b) 30 AMIP models (orange for ensemble mean, grey for individual members), (c-d) ACE2-ERA5 (red for ensemble mean, grey for individual members), and (e-f) NeuralGCM (blue for ensemble mean, grey for individual members).}
    \label{fig:Sup. Figure 5}
\end{figure}

\begin{figure}
    \centering
    \includegraphics[width=\textwidth]{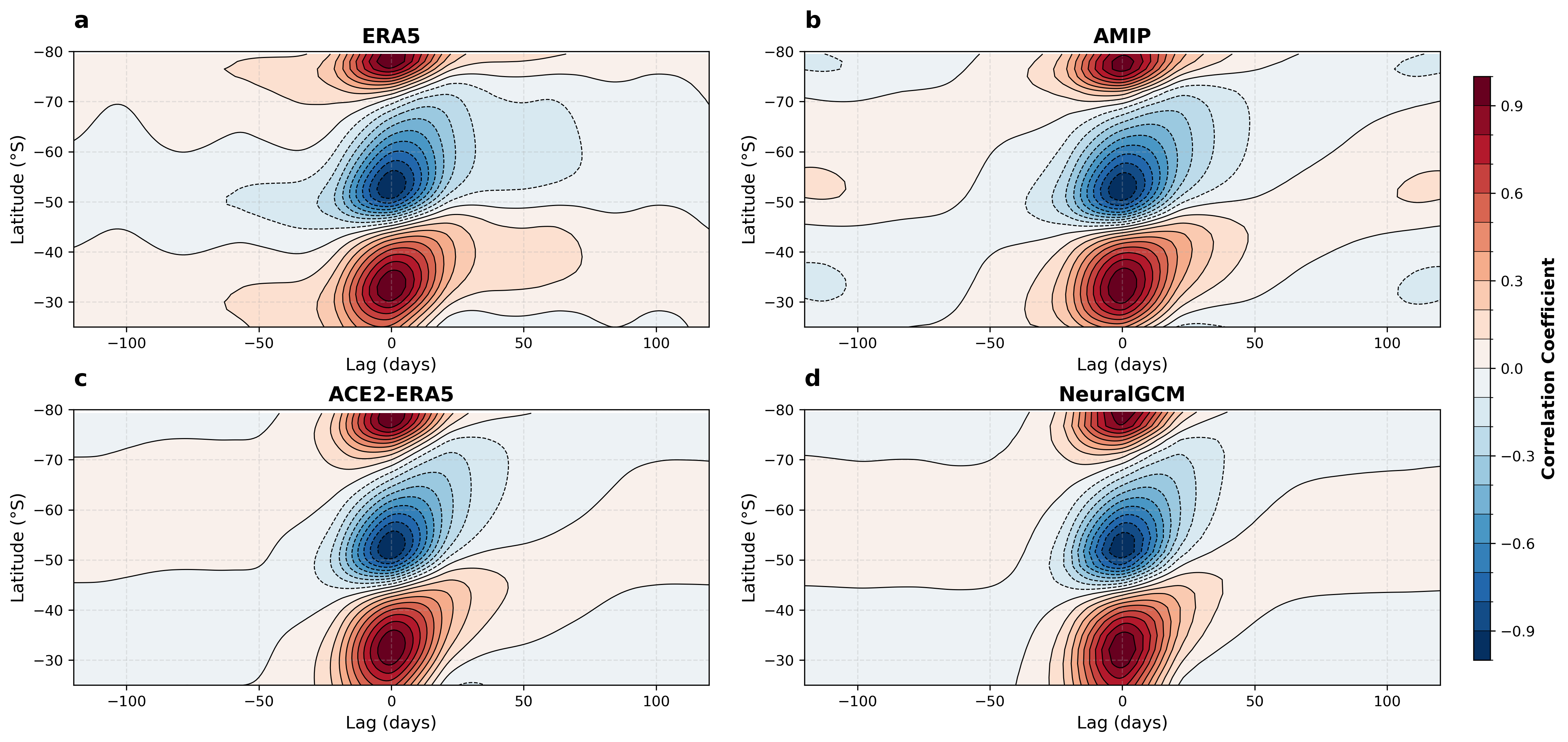}
    \caption{One-point lag-correlation map of the zonal-mean zonal wind anomalies integrated across the depth of the troposphere (1000–100 hPa) for ERA5 and at 500 hPa for AMIP, ACE2-ERA5, and NeuralGCM, reconstructed from projections onto the two leading EOFs of the zonal-mean zonal wind anomalies. The base latitude is at 32.5°S. Contour lines have intervals of 0.1, where the dashed lines indicate negative values.}
    \label{fig:Sup. Figure 6}
\end{figure}

\clearpage

\end{document}